\documentclass[11pt,twoside]{article}
\usepackage{vmargin}
\normalsize
\setpapersize{A4}
\setmarginsrb{3cm}{2.5cm}{3cm}{2.5cm}%
                {\baselineskip}{\baselineskip}{\baselineskip}{2\baselineskip}
\usepackage{xspace}
\usepackage{amsmath}
\usepackage{amscd}
\usepackage{amssymb}
\usepackage{array}
\usepackage{theorem}

\newcommand{\sayTheorem}{Theorem}
\newcommand{\saylemma}{lemma}
\newcommand{\sayLemma}{Lemma}

\newcommand{\sayProposition}{Proposition}
\theoremheaderfont{\normalfont\rmfamily\bfseries\upshape}
\theorembodyfont{\normalfont\rmfamily\upshape}
\theoremstyle{plain}
\newtheorem{Theorem}{\sayTheorem}
\newtheorem{Lemma}[Theorem]{\sayLemma}
\newtheorem{Proposition}[Theorem]{\sayProposition}
\newenvironment{Proof}{\vspace{-\theorempostskipamount}\noindent%
\normalfont\rmfamily\bfseries\upshape Proof\ %
\normalfont\rmfamily\upshape%
}{\hspace*{\fill}%\rule{0.5em}{0.5em}
\mbox{$\blacksquare$}\\[\theorempostskipamount]}

\newcommand{\glick}{\hspace*{\fill}%\framebox[0.5em][c]{\rule{0pt}{0pt}}
\mbox{$\square$}}
\usepackage{robx}
\newlength{\normalparindent}
\newlength{\normalparskip}
\newlength{\newparindent}
\newlength{\newparskip}
\newcommand{\Section}[1]{\section{#1}\setcounter{equation}{0}}

\newcommand{\ordlabel}[1]{\label{#1}}
\newcommand{\MarkovL}[2]{\Tpart_{\Gamma}^{[#1;#2]}}
\newcommand{\mathlabel}[1]{\label{#1}}
\newcommand{\TR}[2]{\sideset{}{^{(#2)}_{#1}}\Tr}
\newcommand{\xbibitem}[1]{\bibitem{#1}}
\newcommand{\Rref}[4]{\xbibitem{#1}%
                        #2,\ %                  % Authors
                        { #4}.%                 % Paper/Journal
}
\newcommand{\RXref}[4]{\xbibitem{#1}%
                        #2,\ %                  % Authors
                        {{``\ignorespaces#3\unskip''}},\ %      % Title
                        { #4}.%                 % Paper/Journal
}
\newcommand{\Publisher}[2]{{\slshape #1 (#2)}}
\newcommand{\IN}[5][]{{\slshape{}in #1 {``\ignorespaces#3\unskip''}, \ed by #2.} \Publisher{#4}{#5}}

\newcommand{\STANDARD}[5][]{\Remph{{#2}{\bfseries #1#3} (#4), #5}}%
\newcommand{\NPB}[3]{\STANDARD[B]{Nucl.\ Phys.\ }{#1}{#2}{#3}}
\newcommand{\IJMP}[3]{\STANDARD{Int.\ J.\ Mod.\ Phys.\ }{#1}{#2}{#3}}
\newcommand{\JSP}[3]{\STANDARD{J.\ Stat.\ Phys.\ }{#1}{#2}{#3}}

\newcommand{\PNAS}[3]{\STANDARD{Proc.\ Natl.\ Acad.\ Sci.\ USA\ }{#1}{#2}{#3}}
\newcommand{\PJM}[3]{\STANDARD{Pac.\ J.\ Math.\ }{#1}{#2}{#3}}
\newcommand{\AJM}[3]{\STANDARD{Am.\ J.\ Math.\ }{#1}{#2}{#3}}
\newcommand{\TAMS}[3]{\STANDARD{Trans.\ Amer.\ Math.\ Soc.\ }{#1}{#2}{#3}} 
\newcommand{\IM}[3]{\STANDARD{Invent.\ Math.\ }{#1}{#2}{#3}}
\newcommand{\JA}[3]{\STANDARD{J.\ Algebra }{#1}{#2}{#3}}

\newcommand{\CMP}[3]{\STANDARD{Commun.\ Math.\ Phys.\ }{#1}{#2}{#3}}

\newcommand{\SMF}[2]{\Remph{Soci\'{e}t\'{e} Math\'{e}matique de France %
Ast\'{e}risque, hors s\'{e}rie,\ (#1), #2}}%
\newcommand{\JPA}[3]{\STANDARD{J.\ Phys.\ {\bfseries A}: Math.\ Gen.\ }{#1}{#2}{#3}}
\newcommand{\horiz}[1]{\overline{#1}}           % horizontal part
\newcommand{\horobj}[1]{\overline{#1}}
\newcommand{\affine}[1]{\widehat{#1}}           % make affine absolutely.
\newcommand{\affobj}[1]{#1}                     % use for roots, weights etc.
\newcommand{\co}[1]{{#1}^{\vee}}
\newcommand{\Weyl}{\horobj{W}}                          % Weyl group
\newcommand{\aWeyl}{\affine{W}}                         % affine Weyl group
\newcommand{\Remph}[1]{\textsl{#1}}
\newcommand{\G}{\mathcal{G}}
\newcommand{\expts}[1]{\mathrm{v}^{\ast}(#1)}
\newcommand{\preword}[1]{\makebox[0pt][r]{\text{#1}\rule{1em}{0pt}}}
\newcommand{\M}{\;}
\newcommand{\dd}{\mathrm{d}}

\newcommand{\Tr}{\operatornamewithlimits{Tr}}

\newcommand{\thmpara}{}

\newcommand{\ie}{i.e.\@\xspace}

\newcommand{\wrt}{w.r.t.\@\xspace}
\newcommand{\phd}{Ph.D.\@\xspace}

\newcommand{\ed}{ed.\@\xspace}
\newcommand{\II}{\hbox{{\rm l{\hbox to 1.5pt{\hss\rm l}}}}}

\newcommand{\RR}{\mathbb{R}}

\newcommand{\ZZ}{\mathbb{Z}}
\newcommand{\Rvec}[1]{{\protect\underset{\protect\smash{\sim}}{\protect\smash[b]{#1}}}}

\newcommand{\Tpart}{\mathcal{Z}}                % Partition Function
\newcommand{\Nlpart}[3]{\Tpart^{(#1;#2,#3)}_{\Gamma}}
\newcommand{\Ncpart}[3]{\Tpart^{(#1;#2,#3)}(q)}

\newcommand{\Nmodel}[3]{(#1;#2,#3)}

\newcommand{\Tsss}{\scriptscriptstyle}
\newcommand{\definedas}{\stackrel{\Tsss\text{def}}{=}}
\newcommand{\ket}[1]{\left| #1 \right\rangle}
\newcommand{\bra}[1]{\left\langle #1 \right|}

\newcommand{\commut}[2]{[#1\,,\,#2]}
\newcommand{\Rset}[3][]{\left\{\:#2\:#1|\:#3\:\right\}}
\newcommand{\TaD}{\widehat{D}}                  % Affine D
\newcommand{\TaA}{\widehat{A}}                  % Affine A
\newcommand{\TaE}{\widehat{E}}                  % Affine E
\newcommand{\TaX}{\widehat{X}}

\newcommand{\Tsikc}[2]{\Omega^{(#1)}_{#2}} % Semi-infinite Kronecker comb.

\newcommand{\SU}[1][2]{\mathsf{SU}(#1)}

\newcommand{\reflection}[1]{\sigma_{#1}}

\newcommand{\coxA}{\affobj{\omega}}
\newcommand{\coxB}{\horiz{\omega}}
\newcommand{\coxE}{\check{\omega}}
\newcommand{\coxEo}{\check{\omega}_1}
\newcommand{\coxEt}{\check{\omega}_2}
\newcommand{\coxAo}{\affobj{\omega}_1}
\newcommand{\coxAt}{\affobj{\omega}_2}

\newcommand{\orbA}[1][]{\affobj{\phi}_{#1}}
\newcommand{\coorbA}[1][]{\affobj{\phi}_{#1}^{\vee}}
\newcommand{\orbB}[1][]{\horiz{\phi}_{#1}}
\newcommand{\coorbB}[1][]{\horiz{\phi}_{#1}^{\vee}}
\newcommand{\gengroup}[1]{\langle #1 \rangle}
\newcommand{\rootSPA}{\affine{\mathsf{h}}^{\ast}}

\newcommand{\weightSPA}{\affine{\mathsf{h}}}

\newcommand{\blf}[2]{\left(#1\,,\,#2\right)}                    % dot product
\newcommand{\ablf}[2]{\left\langle#1\,,\,#2\right\rangle}       % affine dot product
\newcommand{\translat}{\mathsf{t}}
\newcommand{\itranslat}{\mathsf{t}^{-1}}
\newcommand{\liealg}[1][g]{\horobj{\mathsf{#1}}}

\newcommand{\kmalg}[1][g]{\affine{\mathsf{#1}}}
\newcommand{\rootsymb}{\alpha}
\newcommand{\rootB}[1][]{\horiz{\rootsymb}_{#1}}

\newcommand{\rootA}[1][]{\affobj{\rootsymb}_{#1}}
\newcommand{\highroot}{\horiz{\theta}}
\newcommand{\corootA}[1][]{\affobj{\rootsymb}_{#1}^{\vee}}

\newcommand{\rootsyssymb}{\Phi}

\newcommand{\rootSB}[1][]{\horiz{\rootsyssymb}^{#1}}

\newcommand{\rootSA}[1][]{\affine{\rootsyssymb}^{#1}}
\newcommand{\weightsymb}{\rho}
\newcommand{\weightA}[1][]{\affobj{\weightsymb}_{#1}}
\newcommand{\weightB}[1][]{\horiz{\weightsymb}_{#1}}
\newcommand{\lambdaA}{\affobj{\lambda}}
\newcommand{\lambdaB}{\horiz{\lambda}}
\newcommand{\kmvec}[3]{\left( \, #1 \: , \: #2 \: , \: #3 \, \right)}
\newcommand{\restrictmap}[2]{\left.#1\right|_{#2}}

\newcommand{\chebychev}[3][]{\mathcal{U}^{(#2)}_{#1}(#3)}
\newcommand{\dorey}{v}
\newcommand{\irootA}[2][0]{\alpha^{(#1)}_{#2}}

\newcommand{\parity}[1]{P(#1)}
\newcommand{\coxAR}[1][]{\omega_{#1}}

\newcommand{\KosPiA}[1][]{\affine{\Pi}_{#1}}
\newcommand{\KosPiB}[1][]{\horiz{\Pi}_{#1}}
\newcommand{\Rspan}{\operatorname{span}}
\newcommand{\orbSA}{\affine{\mathsf{R}}}
\newcommand{\zroot}[1][]{\smash{\alpha^{(#1)}}}
\newcommand{\zrootX}{\alpha}
\newcommand{\zweight}[1][]{\smash{\rho^{(#1)}}}
\newcommand{\zweightX}{\rho}
\newcommand{\zproj}{\mathcal{J}}
\newcommand{\ldsq}{[\kern-0.15em [}
\newcommand{\rdsq}{]\kern-0.15em ]}
\newcommand{\incase}[3][2]{\ldsq #3 \rdsq_{{}_{#2}}}
\newcommand{\cozroot}[1][]{\smash{\alpha^{(#1)\vee}}}

\newcommand{\half}{\tfrac{1}{2}}
\newcommand{\DD}{\mathrm{D}}
\newcommand{\phz}{\smash[b]{\varphi_{\substack{\rule{0pt}{0.5ex}\\\kern-0.3em 0}}}\kern-0.05em}
\newcommand{\phc}[1][]{\smash[b]{\varphi^{#1}_{\substack{\rule{0pt}{0.5ex}\\\kern-0.3em\text{1}}}}\kern-0.05em}

\newcommand{\bdm}{\begin{displaymath}}\newcommand{\edm}{\end{displaymath}}

\newcommand{\etal}{\textsl{et al}.\@\xspace}

\newcommand{\naively}{na\"{\i}vely\xspace}

\begin{document}
\begin{titlepage}
\vskip 0.5cm
\begin{flushright}
        DTP/99/79 \\
        November 1999 \\
        hep-th/9911058
\end{flushright}
\vskip 1.5cm
\begin{center}
\mathversion{bold}
        \Large\bfseries%
	  Pasquier Models at~$c=1$:\\[5pt]
        Cylinder Partition Functions\\[5pt]
        and the role of the\\[5pt]
        Affine Coxeter Element\\
\end{center}
\vskip 0.9cm
\centerline{Robert P. T. Talbot}
\vskip 0.6cm
\centerline{\slshape Department of Mathematical Sciences,}
\centerline{\slshape University of Durham, Durham DH1 3LE,
England\,\footnote{E-mail addresses: \texttt{r.p.t.talbot@durham.ac.uk} and/or
\texttt{r.p.talbot@gmx.net} (permanent)}}
\vskip 1cm
\begin{abstract}
\vskip0.2cm
	\noindent We calculate the partition functions of the affine Pasquier models on the cylinder in
	the continuum limit.
	We show that the partition function of any affine model may be expressed in terms of the orbit structure
	of the affine Coxeter element of the Weyl group associated with the defining graph of the model.
	Some of the consequences of this geometric relationship are explored.
\bigskip
 
%\noindent PACS numbers: xxxxxxx
\end{abstract}
\end{titlepage}
%
%
%
%%%%%%%%%%%%%%%%%%%%%%%%%%%%%%%%%%%%%%%

%%%%%%%%%%%%%%%%%%%%%%%%%%%%%%%%%%%%%%%

\setcounter{footnote}{0}
\def\thefootnote{\arabic{footnote}}

%%%%%%%%%%%%%%%%%%%%%%%%%%%%%%%%%%%%%%%
%
%
%
%
%
%

%%%%%%%%%%
%
\Section{Introduction and Motivation\ordlabel{sIAM}}
The purpose of this paper is to find a general geometric expression for the partition functions of the
affine Pasquier models on the cylinder.
In so doing we find expressions for the partition functions of the models which are
currently missing from the literature.
In this section we present the problem and take the opportunity to introduce some notation.

The Pasquier models~\cite{pPas87b} are particular examples of graph-lattice models.
They are defined on the square lattice with a height or spin function~$\sigma$ assigning to each lattice vertex
values within the set of nodes of a graph~$\G$.
The values of the heights of two neighbouring vertices are restricted so that 
they are adjacent nodes of~$\G$. This restriction is encoded by the adjacency matrix of the graph
which we shall write as~$\G_{ab}$.
The affine Pasquier models have graphs restricted to the set given in figure~\ref{fAff}.
\begin{figure}[htbp]
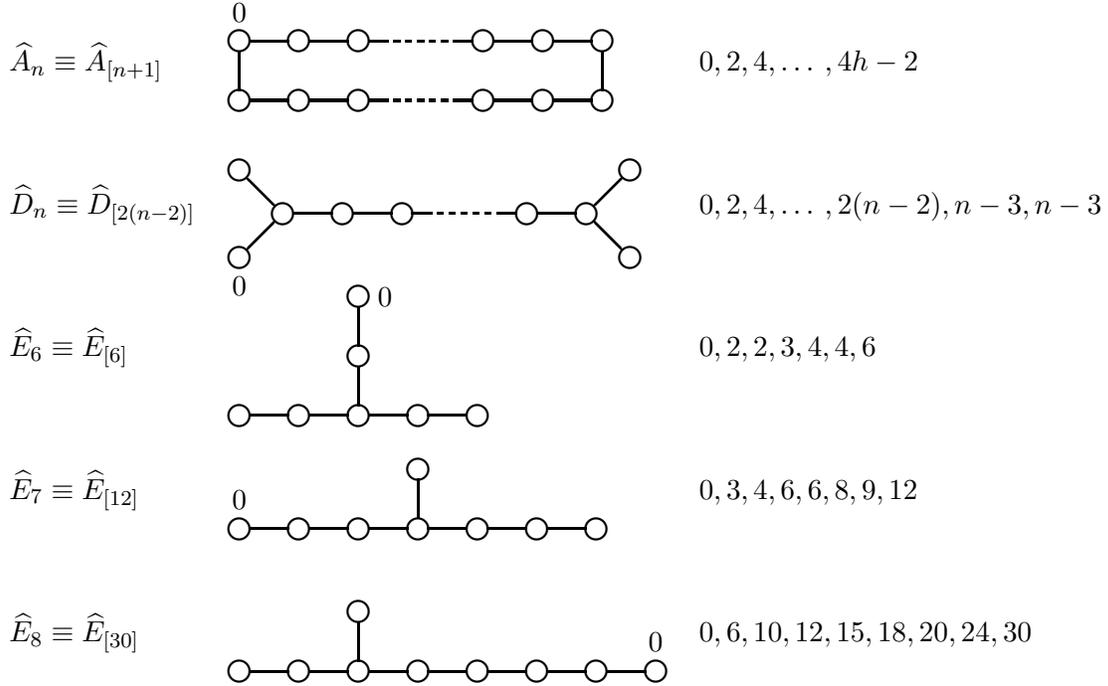

\begin{center}
\begin{tabular}{lll}
{$\TaA_{n}\equiv\TaA_{[n+1]}$}
&
\littlecoxeter\begin{coxeter}{0}{0}{183}{30}{54pt}
\xwb \xe \xwb \xe \xwb \xde \xwb \xe \xwb \xe \xwb \xn \xwb \xw \xwb \xw
\xwb \xdw \xwb \xw \xwb \xw \xwb \xln{$0$} \xs
\end{coxeter}
&
$0,2,4,\ldots,4h-2$
\\
{$\TaD_{n}\equiv\TaD_{[2(n-2)]}$}
&
\littlecoxeter\begin{coxeter}{0}{0}{197}{44}{54pt}
\xls{$0$}
\xwb \xne \xwb \xnw \xwb \xtse \xe \xwb \xe \xwb \xde \xwb \xe \xwb \xse
\xwb \xtnw \xne \xwb
\end{coxeter}
&
$0,2,4,\ldots,2(n-2),n-3,n-3$
\\
{$\TaE_{6}\equiv\TaE_{[6]}$}
&
\littlecoxeter\begin{coxeter}{0}{0}{120}{60}{54pt}
\xwb \xe \xwb \xe \xwb \xn \xwb \xn \xwb \xle{$0$} \xts \xts \xe \xwb \xe \xwb
\end{coxeter}
&
$0,2,2,3,4,4,6$
\\
{$\TaE_{7}\equiv\TaE_{[12]}$}
&
\littlecoxeter\begin{coxeter}{0}{0}{180}{30}{54pt}
\xwb \xln{$0$} \xe \xwb \xe \xwb \xe \xwb \xn \xwb \xts \xe \xwb \xe \xwb \xe \xwb
\end{coxeter}
&
$0,3,4,6,6,8,9,12$
\\
{$\TaE_{8}\equiv\TaE_{[30]}$}
&
\littlecoxeter\begin{coxeter}{0}{0}{210}{30}{54pt}
\xwb \xe \xwb \xe \xwb \xn \xwb \xts \xe \xwb \xe \xwb \xe \xwb \xe \xwb
\xe \xwb \xln{$0$}
\end{coxeter}
&
$0,6,10,12,15,18,20,24,30$
\\
%% Extra diagrams
%{$\TaA_{2n-1}/\ZZ_2$}
%&
%\littlecoxeter\begin{coxeter}{0}{0}{208}{0}{54pt}
%\SLWskip \xselflinkw \xwb \xe \xwb \xe \xwb \xde \xwb \xe \xwb \xe \xwb \xselflinke
%\end{coxeter}
%\\
%{$\TaD_{2n-1}/\ZZ_2$}
%&
%\littlecoxeter\begin{coxeter}{0}{0}{175}{44}{54pt}
%\xwb\xne\xwb\xnw\xwb\xtse \xwb \xe \xwb \xe \xwb \xde \xwb \xe \xwb \xselflinke
%\end{coxeter}
%\\
\end{tabular}
\end{center}
\caption[The affine $\TaA\TaD\TaE$ graphs.]{\ordlabel{fAff}%
The affine graphs and their exponents. Each graph~$\TaX_n$ has~$n+1$ nodes.}
\end{figure}
These graphs (or to be more precise, their adjacency matrices)
are known to have highest eigenvalue~$\beta\equiv\beta^{(0)}=2$. The models they provide are believed
to be integrable and, at criticality, provide conformal theories with
central charge~$c=1$ (see references~\cite{pPasX} and~\cite{pGin88}).
%They arise in the classification of Kac-Moody algebras and
%are generalisations of the~$ADE$ graphs which appear in the classification of simple Lie algebras.
%Indeed, by removing the node labelled~$0$ from a graph~$\TaX_n$ in figure~\ref{fAff}, one obtains
%the Dynkin diagram~$X_n$ of the so-called horizontal subalgebra.
%
These graphs also arise in the classification of affine Kac-Moody algebras just as the classical (non-affine)~$ADE$
Dynkin diagrams appear within the classification of Lie algebras. In fact each affine Kac-Moody algebra~$\kmalg$
is related to an underlying Lie algebra~$\liealg$ known as its horizontal subalgebra. This relationship
is reflected in the diagrams. Removing the ``affine node'' labelled~$0$ (see table~\ref{fAff}) from the
graph~$\TaX_n$ (say) associated with the affine algebra~$\kmalg$, one obtains the graph~$X_n$ of the
horizontal subalgebra~$\liealg$.

We will examine models with a cylindrical rather than toroidal geometry.
If opposing edges of a rectangular~$M\times L$ lattice are identified so that the temporal direction is periodic
then the partition function of the resultant model may be written in the continuum limit as~\cite{pCardy}
\begin{equation}
	\Ncpart{\G}{a}{b}=
	\sum_{n\in\mathcal{J}}
	\Lambda_{n}^{\Nmodel{\G}{a}{b}}\,\chi_{\frac{n^2}{4}}(q)
	\M;
\mathlabel{qCharCoef}
\end{equation}
where the~$\smash{\chi_{\frac{n^2}{4}}(q)}$ are~$c=1$ Virasoro characters; and~$\mathcal{J}$ is some index set labelling the
representations of the single Virasoro algebra present in the theory.
Note that to specify a model one must state the graph which defines its bulk behaviour and the boundary
conditions on either end of the cylinder. Therefore we denote a model
by~$\Nmodel{\G}{a}{b}$. In this paper we are only interested in the simple boundary conditions where the
height function~$\sigma$ is held to constant values~$a,b\in\G$ on each end of the cylinder.
We will denote by~$\Gamma$ the discrete lattice on the cylinder.

The role played by the classical (\ie non-affine) Coxeter element in the partition functions of the
classical Pasquier models was first demonstrated by Dorey~\cite{pDor93}.
The Coxeter element is any one of a class of special elements within the Weyl group of reflections defined
by the geometry of the simple roots.
A Coxeter element is defined as the product of all the reflections associated with the simple
roots (see~\cite{bHumphreys} for a good overview).
In practice we will choose a particular Coxeter element.
If we exclude the graphs of $\TaA$-type with odd numbers of nodes from our analysis,
it follows from the $\ZZ_2$-colourability of the remaining affine graphs
that the roots of an affine Kac-Moody algebra may be disjointly partitioned into two sets of mutually orthogonal
roots~$\Pi_1$ and~$\Pi_2$.
We fix notation by requiring $\rootA[0]$ to be orthogonal to the roots of $\Pi_2$.
Define the involutions~$\coxAo$ and~$\coxAt$ as the products of the reflections associated to the
sets~$\Pi_1$ and~$\Pi_2$ of simple roots respectively.
The exact ordering of the roots within either of these sets is irrelevant as they are mutually orthogonal.
Then the Steinberg-ordered Coxeter element~\cite{pSte59} is defined as
\begin{equation}
	\coxA\definedas\coxAt\,\coxAo\M.
	\mathlabel{qCoxChoice}
\end{equation}
It is this particular choice of Coxeter element that we will find easiest to manipulate.

The generalisation to the affine Coxeter element is non-trivial however.
The order of (any of)
the Coxeter elements is known as the Coxeter number. In the affine cases, this order is infinite and an
analogous number known as the affine Coxeter number~$h$ (finite) takes its place in expressions for eigenvalues
and eigenvectors of the adjacency matrix. (We will often find it convenient to label graphs and their
eigenvectors and eigenvalues with this number in which case we will do so by placing square parentheses around
the Coxeter number, $\G_{[h]}$ for example.)
The affine Weyl group~$\aWeyl$ contains translations as well as ordinary reflections.
We shall see that the action of the Affine Coxeter element can be thought of as the action of something not
unlike the action of the Coxeter element in the root space of the underlying horizontal algebra (this
action has finite order~$h$) combined with the additional effect produced by a translation.
There are also ``imaginary'' roots, pointwise invariant under~$\aWeyl$ complicating the geometry.
A good overview of the relationship between affine Weyl groups and
Kac-Moody algebras may be found in~\cite{bFuchs}.
%Good overviews of the affine Weyl group and its relationship to Kac-Moody algebras may be found
%in~\cite{bHumphreys} and~\cite{bFuchs} respectively.

%
Dorey~\cite{pDor93} demonstrated that the coefficients~$\Lambda^{\Nmodel{\G}{a}{b}}_n$ in the expression
analogous to~\eqref{qCharCoef} for the non-affine models ($c<1$) had a simple geometric interpretation in terms of
the orbit structure of the (non-affine analogue of the) Coxeter element~\eqref{qCoxChoice}.
We extend this result to the affine models ($c=1$) and in so doing also provide an alternative method
of proof of the result for the non-affine cases.
We will also be interested in the geometric consequences of the relationship mentioned earlier
between an affine Kac-Moody algebra and its underlying Lie algebra.

We begin in the next section by calculating the coefficients~$\Lambda_{n}^{\Nmodel{\TaA_{[2h]}}{a}{b}}$
for the model $\Nmodel{\TaA_{[2h]}}{a}{b}$.
In section~\ref{sIntertwiners} we construct affine generalisations of intertwiners which
will relate the remaining $\TaA$-, $\TaD$- and $\TaE$-based models to this basic model.
We show that all the partition functions have a single algebraic form in section~\ref{sASFFTPF}.
In section~\ref{sTROCG}, we rewrite this form in terms of the geometry of the Weyl group of
reflections and finally, in section~\ref{sGC} we explore some of the consequences of this geometric
interpretation.

%%%%
%
\Section{The Partition Function of the Model~$\Nmodel{\TaA_n}{a}{b}$\ordlabel{sTPFOTAM}}
We consider first the model~$\Nmodel{A_{\infty}}{a}{b}$.
The height function, $\sigma$, of this model is
the discrete version of a continuous free field variable~$\varphi\in\RR$.
Up to rescaling of~$\varphi$, this theory has partition function
\begin{equation}
        \Tpart^{\Nmodel{A_{\infty}}{a}{b}}(M,L)=\rule{-2em}{0pt}
                \operatornamewithlimits{\mbox{$\displaystyle\int$}}_{\substack{\varphi(a^0,0)=a\\\varphi(a^0,L)=b\\
                                                                \varphi(a^0+M,a^1)=\varphi(a^0,a^1)}}
                \rule{-2em}{0pt}\DD\varphi\:\exp\left\{-\frac{\pi g}{4}\int\dd^2 a\:
                \big|\nabla\varphi(a^0,a^1)\big|^2\right\}
        \M,
\mathlabel{qGaussian}
\end{equation}
where~$g=1$~\cite{pSaleur}. This model is well known in the literature to provide a~$c=1$ conformal field %88
theory (see~\cite{lGinACFT} for example).
The equivalence,
\begin{equation}
        \lim_{\substack{d\rightarrow 0\\M,L\rightarrow\infty\\\text{$M/L$ fixed}}}
        \Nlpart{A_{\infty}}{a}{b}=\Tpart^{\Nmodel{A_{\infty}}{a}{b}}(M,L)
        \M,
\end{equation}
of the lattice model in the continuum limit to the Gaussian or
Coulomb-gas model~\eqref{qGaussian}
is established by examining the renormalisation-group flows of the so-called (unrestricted)
solid-on-solid model mapped onto the~$A_{\infty}$ model via a set of transformations~\cite{pNie84}.

The partition function~\eqref{qGaussian} is calculated by
splitting~$\varphi(a^0,a^1)=\phz(a^0,a^1)+\phc(a^1)$.
The field~$\phc(a^1)$ is the classical solution to the equations of motion.
The field~$\phz$ is periodic in the time direction
and satisfies the boundary conditions~$\phz(a^0,0)=\phz(a^0,L)=0$.
By zeta-regularisation~\cite{pSaleur}, %87
\begin{equation}
        \operatornamewithlimits{\mbox{$\displaystyle\int$}}_{\substack{\phz(a^0,0)=\phz(a^0,L)=0\\
                \rule{0pt}{1ex}\\\phz(a^0+M,a^1)=\phz(a^0,a^1)\\\rule{0pt}{1ex}}}
                \rule{-2em}{0pt}\DD\phz\:\exp\left\{-\frac{\pi g}{4}\int\dd^2 a\:
                \big|\nabla\phz\big|^2\right\}
        =\eta^{-1}(q)
        \M,
\mathlabel{qZeta}
\end{equation}
with~$q\equiv\exp\left(-M\pi/L\right)$, the modular parameter.

The classical field~$\phc(a^1)=(b-a)\,a^1/L$ \cite{pSaleur}, so~\eqref{qGaussian} is %88
\begin{equation}
        \Tpart^{\Nmodel{A_{\infty}}{a}{b}}(q)=\eta^{-1}(q)\,\exp\left\{
                -\frac{\pi g}{4}\,\left(\frac{b-a}{L}\right)^2 ML \right\}
        =\eta^{-1}(q)\,q^{g\,(b-a)^2/4}
        \M.
\end{equation}
The model is (manifestly) invariant under shifts in the boundary conditions~$(a,b)\rightarrow (a+c,b+c)$, so we
set~$\varepsilon\equiv b-a$; thus, setting also~$g=1$,
\begin{equation}
%\Rboxed{
        \Tpart^{\Nmodel{A_{\infty}}{0}{\varepsilon}}(q)=\eta^{-1}(q)\,q^{\varepsilon^2/4}
%}
        \M.
\mathlabel{qSaleur}
\end{equation}

By graph-symmetry, the partition function of the model~$\Nmodel{\TaA_{2h-1}}{a}{b}$ is also invariant under
`translation' of the boundary conditions. We again rewrite the boundary condition as~$\varepsilon\equiv b-a$.
The continuum limit of the partition function is found by noting that the height function~$\sigma$ is here the
discrete version of the continuous field~$\varphi\in\RR/2h\ZZ$; \ie the field~$\varphi$ of~\eqref{qGaussian}
is now compactified on a circle of circumference~$2h$~\cite{pdiF92}. With a suitable rescaling of the
field~$\varphi$, the continuum partition function will be given by~\eqref{qGaussian}, again with~$g=1$.
The partition function~\eqref{qGaussian} is evaluated by again splitting~$\varphi(a^0,a^1)=\phz(a^0,a^1)+\phc(a^1)$.
The field~$\phz(a^0,a^1)$ is as before and provides the contribution~\eqref{qZeta}. The classical
field~$\phc(a^1)$ is modified to identify the boundary condition~$\varepsilon$ with~$\varepsilon+2nh$
for~$n\in\ZZ$. Thus
\begin{equation}
        \phc(a^1)\equiv\sum_{n=-\infty}^{\infty}\phc[(n)](a^1)
        \M;
\end{equation}
where~$\phc[(n)](a^1)$ is the classical field in the~$n^{\text{th}}$ instanton sector and is
\begin{equation}
        \phc[(n)](a^1)=(\varepsilon+2nh)\,\frac{a^1}{L}
        \M.
\end{equation}
Thus the continuum partition function
of the model~$\Nmodel{\TaA_{2h-1}}{0}{\varepsilon}$ is given by the result~\eqref{qSaleur} with the
identification~$\varepsilon\equiv\varepsilon+2\,h$, \ie
\begin{equation}
%\Rboxed{
        \Tpart^{\Nmodel{\TaA_{2h-1}}{0}{\varepsilon}}(q)=\eta^{-1}(q)\,\sum_{n\in\ZZ}
                q^{\frac{(\varepsilon+2nh)^2}{4}}
%}
        \M.
\mathlabel{qSaleurGen}
\end{equation}

The partition functions~\eqref{qSaleur} and~\eqref{qSaleurGen} can be placed in the form~\eqref{qCharCoef}
by noting that
\begin{equation}
	\eta^{-1}(q)\;q^{\frac{n^2}{4}}=
	\sum_{p=0}^{\infty}
	\chi_{\frac{(n+2p)^2}{4}}(q)
	\mathlabel{qInvert}
\end{equation}
where~$n$ is a positive integer and~$\smash{\chi_{\frac{n^2}{4}}(q)}$ are
\Remph{degenerate}~$c=1$ characters (refer to references~\cite{pdiFSZ87}
and~\cite{aKac79} for details).
If we define the \Remph{generalised semi-infinite Kronecker comb} as
\begin{equation}
	\Tsikc{a_{1},\dots,a_{k}}{n,\varepsilon}
	\equiv
	\sum_{m_{1},\dots,m_{k}=0}^{\infty}
	\delta_{n,\varepsilon+\smash{\underset{i=1}{\overset{k}{\sum}}}m_{i}\,a_{i}}
	\M,
\mathlabel{qSIKC}
\end{equation}
where~$\delta$ is the usual Kronecker-$\delta$; then~\eqref{qSaleur}
is of the form~\eqref{qCharCoef} with
\begin{equation}
	\Lambda_{n}^{\Nmodel{A_{\infty}}{0}{\varepsilon}}=
	\Tsikc{2}{n,\varepsilon}
	\M.
\mathlabel{qiApfC}
\end{equation}
Similarly, defining~$\varepsilon\equiv\varepsilon+2nh$,
it is an easy exercise to see that~\eqref{qSaleurGen} is of the form~\eqref{qCharCoef} with
\begin{equation}
	\Lambda^{\Nmodel{\TaA_{[2h]}}{0}{\varepsilon}}_{n}=
	\Tsikc{2,2h}{n,2h-\varepsilon}+\Tsikc{2,2h}{n,\varepsilon}
	\M.
\mathlabel{qaApfC}
\end{equation}
We could deduce the partition functions of the models based upon the odd cycles, $\TaA_{[2h-1]}$, by
substituting~$2h\mapsto2h-1$. However the $\TaA_{[2h]}$-based models form a kind of ``basis set''
as we shall see below.

%%%%
%
\Section{Intertwiners Between Affine Models\ordlabel{sIntertwiners}}
We now have an expression~\eqref{qaApfC} for the partition function of the models
of the form $\Nmodel{\TaA_{[2h]}}{0}{\varepsilon}$. We construct the partition functions
of~$\TaD$- and $\TaE$-based models and of the odd cycle~$\TaA$ models
by re-expressing these models, still on the finite lattice~$\Gamma$,
in terms of partition functions of $\TaA_{[2h]}$-based models. In other words, we construct the
intertwiners interrelating the models. These relationships between the models continue to hold in
the continuum limit and we then may use~\eqref{qaApfC} to construct explicit partition functions
for the~$\TaA_{[2h-1]}$-, $\TaD$- and $\TaE$-based cases.

We wish to find coefficients~$N^{ab}_{0\lambda ;h'}$ such that
\begin{equation}
        \Nlpart{\G_{[h]}}{a}{b}=
        \sum_{h'}\sum_{\lambda\in\TaA_{[h']}}
        N^{ab}_{0\lambda ;h'}\,
        \Nlpart{\TaA_{[h']}}{0}{\lambda}
        \M.
\mathlabel{qFMdecom}
\end{equation}
Denote by~$\Tr_{(a,b)}$ the trace subject to the boundary conditions~$(a,b)$.
The Markov trace is defined for any operator~$\mathsf{Y}$ on the Hilbert space of a model based on the graph~$\G$ by
\begin{equation}
\begin{split}
	\TR{\sigma_1}{\nu}\mathsf{Y}&\definedas
	\sum_{\sigma_L}
	\frac{\psi_{\sigma_L}^{(\nu)}}{\psi_{\sigma_1}^{(\nu)}}\,
	\Tr_{(\sigma_1,\sigma_L)}\mathsf{Y}\\
	&\equiv
	\sum_{\sigma_2,\ldots,\sigma_L}
	\frac{\psi_{\sigma_L}^{(\nu)}}{\psi_{\sigma_1}^{(\nu)}}
	\bra{\sigma_1,\ldots,\sigma_L}\mathsf{Y}\ket{\sigma_1,\ldots,\sigma_L}
	\M;\\
\end{split}
\mathlabel{qModTrace}
\end{equation}
where the~$\Rset[\big]{\Rvec{\psi}^{(\mu)}}{\mu\in\expts{\G}}$ are the eigenvectors
of the graph~$\G$ labelled by the exponents of~$\G$. 
The modified partition function is defined~\cite{pSoc} in terms of this trace by
\begin{equation}
\begin{split}
	\MarkovL{\G}{\nu}
	&\definedas
	\TR{a}{\nu}T^M
	\\
	&\equiv
	\sum_{b\in\G}\frac{\psi_b^{(\nu)}}{\psi_a^{(\nu)}}\,\Nlpart{\G}{a}{b}
	\M.\\
\end{split}
\mathlabel{qModPartFn}
\end{equation}
Sochen~\cite{pSoc} has demonstrated the remarkable property
that this partition function is both independent of~$a$
and that it is graph independent: any two graphs, $\G_1$ and~$\G_2$, sharing the same exponent~$\nu$ and
highest (Perron-Frobenius) eigenvector~$\beta$ have identical modified partition functions
on the lattice~$\Gamma$, \ie
\begin{equation}
	\MarkovL{\G_1}{\nu}=\MarkovL{\G_2}{\nu}
	\M.
\mathlabel{qSochenLemmaResult}
\end{equation}
The solution to this equation~\eqref{qSochenLemmaResult}, with~$\G_1$ and~$\G_2$ both chosen to be (non-affine)
$ADE$ graphs with equal Coxeter number yields the intertwiners between the~$A$-, $D$- and~$E$-based models.
Although typically no~$\TaA$ graph shares all the exponents of any given~$\TaD$ or~$\TaE$ graph, the
graph independence of~\eqref{qModPartFn} can still be used to interrelate the
affine models as follows:
Let~$\Rset{\beta^{(\mu)}}{\mu\in\expts{\G}}$
%(with $\beta\equiv\beta^{(0)}=2$ the Perron-Frobenius eigenvalue)
denote the eigenvalues of~$\G$ and let
\begin{equation}
	\MarkovL{\G_{[h]}}{\nu}=f(\beta,\beta^{(\nu)})
	\M,
\mathlabel{qSocLemR2}
\end{equation}
where~$f$ is some graph independent function,
denote the modified partition function for a model based on the affine graph~$\G$
with affine Coxeter number~$h$ (see~\cite{pSoc}).
We seek solutions to the more general equation
\begin{equation}
	\MarkovL{\G_{[h]}}{\nu}=
	\sum_{\substack{h'\in\ZZ^{+}\\\nu'\in\TaA_{[h']}}}
	X^{h\nu}_{h'\nu'}\,
	\MarkovL{\TaA_{[h']}}{\nu'}
	\M.
\mathlabel{qMarkovDecom}
\end{equation}
We allow freedom both in the choice of the $\TaA$ graph (labelled by its Coxeter number~$h'$) and in the
exponent~$\nu'$.
Substituting in~\eqref{qSocLemR2} and labelling the eigenvalues of a graph with Coxeter number~$k$
by~$\beta^{(\mu)}_{[k]}$, we therefore seek a solution for the~$X^{h\nu}_{h'\nu'}$ such that,
\begin{equation}
	f(\beta_{[h]},\beta_{[h]}^{(\nu)})=\sum_{\substack{h'\in\ZZ^{+}\\\nu'\in\TaA_{[h']}}}
	X^{h\nu}_{h'\nu'}\,f(\beta_{[h']},\beta_{[h']}^{(\nu')})
	\M.
\mathlabel{qModifiedDecom}
\end{equation}
By property~\eqref{qSochenLemmaResult},
the unknown function~$f$ is identical on each side for any given fixed lattice~$\Gamma$.

Given the eigenvalues~$\beta_{[h]}$ and~$\beta_{[h]}^{\nu}$ on the left hand
side of~\eqref{qModifiedDecom} we can solve the equation by
choosing the eigenvalues~$\{\beta_{[h']}^{(\nu')}\}$ on the
right hand side appropriately.
Examining the general form of the
eigenvalues~$\beta_{[k]}^{(\mu)}=2\cos\frac{\mu\pi}{k}$ one might \naively attempt
to choose~$h'=h$ and~$\nu'=\nu$. However, this would require that the
exponent set of the graph~$\G_{[h]}$ be a subset of the
exponent set of the graph~$\TaA_{[h]}$
(\ie~$\expts{\G_{[h]}}\subseteq\expts{\TaA_{[h]}}$) and, as mentioned already, this is not the
case in general.
(For example, $\TaD_5$ has the odd exponent~$3$ occurring with multiplicity~$2$. However, not only
are there no~$\TaA$-based models with odd exponents, but no exponent of any~$\TaA$ graph has
a multiplicity exceeding~$1$.)
Fortunately we may circumnavigate this inconvenience by choosing instead~$h'=2h$ and~$\nu'=2\nu$, \ie
\begin{equation}
        X^{h\nu}_{h'\nu'}=\delta^{2h}_{h'}\,\delta^{2\nu}_{\nu'}
        \M.
\mathlabel{qASoln}
\end{equation}
This provides a solution to~\eqref{qMarkovDecom}.
%We observe that in contrast to the toroidal case~\cite{pPasX}, we can construct a solution whereby a model of (affine)
%Coxeter number~$h$ is expressed in terms of other models, all of which have the \Remph{same} (affine) Coxeter
%number, namely~$2h$. In the toroidal case, the extra degree of freedom provided by the summation over~$h'$
%in~\eqref{qFMdecom} is necessary.
Substituting~\eqref{qASoln} and~\eqref{qModPartFn} into~\eqref{qMarkovDecom}
we obtain,
\begin{equation}
        \sum_{b\in\G_{[h]}}\frac{\psi_b^{(\nu)}}{\psi_a^{(\nu)}}\,\Nlpart{\G_{[h]}}{a}{b}
        =
        \sum_{d\in\TaA_{[2h]}}\frac{\phi_d^{(2\nu)}}{\phi_c^{(2\nu)}}\,\Nlpart{\TaA_{[2h]}}{c}{d}
        \M;
\end{equation}
where the~$\{\Rvec{\psi}^{(\mu)}\}$ are eigenvectors of~$\G_{[h]}$ and the~$\{\Rvec{\phi}^{(\mu)}\}$ eigenvectors
of~$\TaA_{[2h]}$. Both~$a$ and~$c$ may be chosen arbitrarily on the respective graphs.
Assuming the~$\{\Rvec{\psi}^{(\mu)}\}$ have been chosen to be an \Remph{orthonormal} set,
this is inverted to yield,
\begin{equation}
        \Nlpart{\G_{[h]}}{a}{b}=
        \sum_{\lambda\in\TaA_{[2h]}}
        \left\{
        \sum_{\mu\in\expts{\G}}\frac{\phi^{(2\mu)}_{\lambda}}{\phi^{(2\mu)}_{0}}\,
        \psi^{(\mu)\ast}_{a}\,\psi^{(\mu)}_{b}
        \right\}
        \Nlpart{\TaA_{[2h]}}{0}{\lambda}
        \M;
\mathlabel{qGeneral}
\end{equation}
where we set~$c=0$ (the affine node) without loss of generality.
This is of the form~\eqref{qFMdecom} with
\begin{equation}
        N^{ab}_{0\lambda ;h'}=\delta_{h',2h}\,N^{ab}_{\lambda}
\mathlabel{qCoeffs}
\end{equation}
and
\begin{equation}
%\Rboxed{
        N^{ab}_{\lambda}\equiv
        \sum_{\mu\in\expts{\G}}\frac{\phi^{(2\mu)}_{\lambda}}{\phi^{(2\mu)}_{0}}\,
        \psi^{(\mu)\ast}_{a}\,\psi^{(\mu)}_{b}
%}
        \M.
\mathlabel{qDouble}
\end{equation}
%(Remember that here the~$\Rset[\big]{\Rvec{\psi}^{(\mu)}}{\mu\in\expts{\G}}$
%are understood to form an \Remph{orthonormal} eigenbasis of~$\G$.)
We remark that although complete freedom was permitted in the~$\TaA$ models appearing
in equation~\eqref{qFMdecom},
\eqref{qGeneral}~demonstrates that the even cycles provide a basis
set from which all the other models may be constructed.

Equation~\eqref{qCoeffs}, together with equation~\eqref{qDouble} clearly provide
a solution to the decomposition~\eqref{qFMdecom} but not necessarily the %SPELLING
only one. In general there are a number of different possible solutions.
However, any non-trivial solution is suitable for our purposes and we do not investigate other possibilities
here.

The coefficients~\eqref{qDouble} are the analogues of the classical intertwiners
\begin{equation}
	V^{\lambda}_{ab}\equiv
	\sum_{\mu\in\expts{\G}}\frac{\phi_{\lambda}^{(\mu)}}{\phi_{1}^{(\mu)}}\,
		\psi_{a}^{(\mu)\ast}\,\psi_{b}^{(\mu)}
	\M.
\mathlabel{qSocInt}
\end{equation}
They possess a similar algebraic form and play a similar role.
They differ immediately chiefly in that they relate a model with affine Coxeter number~$h$ to a model with
affine Coxeter number~$2h$ and even interrelate~$\TaA$ models!
In contrast, intertwiners interrelate classical models with the Coxeter number preserved in the
relation.

It is certainly tempting, given the form~$\phi^{(\mu)}_{[h]\varepsilon}=\cos\frac{\mu\pi\varepsilon}{h}$ for the
eigenvectors of the~$\TaA_{[h]}$ graph, to identify
the coefficients~\eqref{qDouble} with the classical intertwiners~\eqref{qSocInt}.
Denote by square-parentheses the Coxeter number of the graph to which a vector refers. Then, na\"{\i}vely
\begin{equation}
	\frac{\phi^{(2\mu)}_{[2h]\lambda}}{\phi^{(2\mu)}_{[2h]0}}
	\equiv
	\frac{\phi^{(\mu)}_{[h]\lambda}}{\phi^{(\mu)}_{[h]0}}
	\M.
\end{equation}
However the terms on the right-hand side of this equality are not defined
unless~$\mu$ is even (\ie an exponent of the~$\TaA$ graph);
the more general summation over the full exponent set of the graph~$\G$ implies that this is not always the case.
Furthermore, upon examination, the coefficients~\eqref{qDouble} appear to be both half-integral and
of either sign; properties not shared by~\eqref{qSocInt}.

We\label{PParity} remark that for the $\TaD$ and $\TaE$ cases,
the coefficient~$N^{ab}_{\lambda}$ of~\eqref{qDouble} will be zero if the 
$\ZZ_2$-parity of the length of the path between~$a$ and~$b$ on~$\G_{[h]}$ is not the same as the parity of the
path between~$0$ and~$\lambda$ on~$\TaA_{[2h]}$.
The argument is quite complicated and we do not repeat it here
(it is detailed in full in~\cite{hTal}).
Thus if the original model~$\G_{[h]}$ has even (odd) boundary conditions (\ie the path length between~$a$
and~$b$ is even (odd)) then the $\TaA$-models onto which it decomposes
will also have strictly even (odd) boundary conditions.

%%%%%%%%%%
%
\Section{A Simple Form for the Partition Functions\ordlabel{sASFFTPF}}
While it is possible at this stage to calculate the intertwiners~\eqref{qDouble} and thereby the
partition functions, we shall instead
show that the partition functions may be unified into a single simple form.
For the explicit calculation of the intertwiners and from these the partition functions,
the interested reader is referred elsewhere~\cite{hTal}.

It has been noted (see for example~\cite{pdiFZ90} or~\cite{pdiF92}) that the intertwiners~\eqref{qSocInt} interrelating the
classical (\ie non-affine) Pasquier models may be written as Chebychev polynomials of the
second kind.
These polynomials~$\Rset{\chebychev{n}{x}}{n\in\ZZ}$ are defined by the recursion relation
\begin{equation}
	\chebychev{n}{x}=x\:\chebychev{n-1}{x}-\chebychev{n-2}{x}
	\M;
\mathlabel{qChebRecurs}
\end{equation}
together with two `initial conditions': the values of two
subsequent polynomials in the series. 
We are concerned only with the series of polynomials generated by the
initial conditions:
\begin{equation}
	\chebychev{0}{x}=1\M, \qquad \chebychev{1}{x}=x \M;
\mathlabel{qChebInit}
\end{equation}
and we note that this implies that $\chebychev{-1}{x}=0$.
The classical intertwiners~\eqref{qSocInt} may be written
\begin{equation}
	V^{\lambda}_{ab}\equiv\chebychev[ab]{\lambda-1}{\G}
	\M.
\mathlabel{qdiFResult}
\end{equation}
We propose the same form for the affine intertwiners~\eqref{qDouble}.
We prove that this is indeed the case now:

\begin{Lemma}
\ordlabel{TAIsCheb}
\ordlabel{TTennessee}
	The coefficient $\Lambda^{\varepsilon}_{n}\equiv\Lambda^{\Nmodel{\TaA_{[2h]}}{0}{\varepsilon}}_{n}$
	of $\chi_{\frac{n^2}{4}}(q)$ in the
	expansion~\eqref{qCharCoef} of the partition function of the
	model~$\Nmodel{\TaA_{[2h]}}{0}{\varepsilon}$ is given by
	\begin{equation}
		\Lambda^{\varepsilon}_{n}=\chebychev[0\varepsilon]{n}{\TaA_{[2h]}}
		\M.
	\mathlabel{qAIsCheb}
	\end{equation}
	In other words, it is given by the $n^{\text{th}}$-Chebychev polynomial of the
	second kind satisfying $\chebychev{0}{\TaA}=\II$ and $\chebychev{1}{\TaA}=\TaA$.
\glick\end{Lemma}
\begin{Proof} The proof is by induction on the variable $n$.

\thmpara
We first establish that the initial conditions \eqref{qChebInit} for the Chebychev recursion relation
are satisfied. We observe, directly from~\eqref{qaApfC}, that
\begin{equation}
	\Lambda^{\varepsilon}_{0}=\delta^{\varepsilon}_{0}=\chebychev[0\varepsilon]{0}{\TaA_{[2h]}}
\end{equation}
and
\begin{equation}
	\Lambda^{\varepsilon}_{1} = \delta^{\varepsilon}_{1} + \delta^{\varepsilon}_{2h-1}
	=\chebychev[0\varepsilon]{1}{\TaA_{[2h]}}
\end{equation}
as required.

\thmpara
Having established that the initial conditions are satisfied, we now examine the
recursion relation~\eqref{qChebRecurs}. Assume that $\exists$~$N\in\ZZ^{+}$ such that $\forall$~$2\leq n<N$
that \eqref{qAIsCheb} holds. Then
\begin{equation}
\begin{split}
	\left[
	\chebychev{N-1}{\TaA_{[2h]}}\:\TaA_{[2h]}
	- \chebychev{N-2}{\TaA_{[2h]}}
	\right]_{0\varepsilon}
	&=
	\sum_{\lambda\in\TaA_{[2h]}} \Lambda^{\lambda}_{N-1}\, 
		\TaA_{\lambda\varepsilon} - \Lambda^{\varepsilon}_{N-2}		\\
	&=
	\Lambda^{\varepsilon-1}_{N-1} + \Lambda^{\varepsilon+1}_{N-1} - \Lambda^{\varepsilon}_{N-2} \\
\end{split}
\mathlabel{qL1T1}
\end{equation}
using the induction hypothesis. Note that, as defined in section~\ref{sTPFOTAM},
the~$\Lambda^{\varepsilon}_n$ have a 
period~$2h$ in the variable $\varepsilon$, so that addition and subtraction in this
variable (such as in~\eqref{qL1T1}) are understood to be taken \Remph{modulo~$2h$}.

The righthand side of~\eqref{qL1T1} will be equal to the required $\Lambda^{\varepsilon}_{N}$
iff
\begin{equation}
	\Lambda^{\varepsilon-1}_{N-1} + \Lambda^{\varepsilon+1}_{N-1}
	=
	\Lambda^{\varepsilon}_{N} + \Lambda^{\varepsilon}_{N-2}
	\M.
\mathlabel{qL1T2}
\end{equation}
This equation is identically true $\forall$ $N$ replaced with $n>2$ as a detailed analysis of the
coefficients $\Lambda^{\varepsilon}_{n}$ shows. If we examine a typical grid
of values $\Lambda^{\varepsilon}_{n}$ (given by equation~\eqref{qaApfC}), as in figure~\ref{tA5Grid}. 
\begin{figure}[t]		% htb
\begin{center}
\setlength{\extrarowheight}{5pt}
\begin{tabular}{|c||cccccc|cccccc|ccc|}
\hline
$\varepsilon$ & \multicolumn{15}{c|}{$n$}\\
\hline
  & {\bf 0} & 1 & 2 & 3 & 4 & 5 & {\bf 6} & 7 & 8 & 9 & 10 & 11 & {\bf 12} & 13 & 14 \\
\hline
\hline
  & &&&&&& &&&&&& &&\\
4 & {\bf 0} &   & 1 &   & {\bf 2} &   & {\bf 2} &   & 3 &   &  {\bf 4} &    & {\bf  4} &    & 5  \\
5 &         & 1 &   & 1 &   & {\bf 2} &         & 3 &   & 3 &    &  {\bf 4} &          &  5 &    \\
\hline
0 & 1 &   & 1 &   & 1 &   &  3 &   & 3 &   &  3 &    & 5 &    & 5  \\
1 &         & 1 &   & 1 &   & {\bf 2} &         & 3 &   & 3 &    &  {\bf 4} &          &  5 &    \\
2 & {\bf 0} &   & 1 &   & {\bf 2} &   & {\bf 2} &   & 3 &   &  {\bf 4} &    & {\bf  4} &    & 5  \\
3 &         & {\bf 0} &   & {\bf 2} &   & {\bf 2} &         & {\bf 2} &   & {\bf 4} &    &  {\bf 4} &  &  {\bf 4} &    \\
4 & {\bf 0} &   & 1 &   & {\bf 2} &   & {\bf 2} &   & 3 &   &  {\bf 4} &    & {\bf  4} &    & 5  \\
5 &         & 1 &   & 1 &   & {\bf 2} &         & 3 &   & 3 &    &  {\bf 4} &          &  5 &    \\
\hline
0 & 1 &   & 1 &   & 1 &   & 3 &   & 3 &   &  3 &    & 5 &    & 5  \\
1 &         & 1 &   & 1 &   & {\bf 2} &         & 3 &   & 3 &    &  {\bf 4} &          &  5 &    \\
  & &&&&&& &&&&&& &&\\
\hline
\end{tabular}
\end{center}
\caption[The coefficients $\Lambda_{n,\varepsilon}$ for the model $\TaA_5$.]{\ordlabel{tA5Grid}%
The coefficients $\Lambda_{n,\varepsilon}$ for the model $\TaA_5$.
The multiples of $n=h\equiv 6$ are highlighted in bold. Due to `parity constraints',
%(see the remark on page~\pageref{PParity})
every second
value is automatically zero and these values have been left blank for clarity.
The values in the figure are `coloured'
alternatively bold and normal to emphasise the ``embedded-diamond''-like pattern of the numbers.
Note that the figure is periodic in the variable~$\varepsilon$ (see text) and that~$n$ takes
values from~$0$ to~$\infty$. This same basic
pattern of numbers occurs for each of the $\ZZ_2$-colourable $\TaA$-based models.}
\end{figure}
Fixing arbitrarily, the values of $\varepsilon$ and $n$ we see that $\Lambda^{\varepsilon}_{n}$
takes the value, \Remph{$m$ say}, with coordinates $\varepsilon$ and $n$ in the grid, 
$\Lambda^{\varepsilon}_{n-2}$ takes the
values two spaces to the left (or `west') of this value, $\Lambda^{\varepsilon-1}_{n-1}$ takes the
value `northwest' and $\Lambda^{\varepsilon+1}_{n-1}$ the value `southwest'. Due to the periodicity
of the variable~$\varepsilon$ and the ``embedded-diamond''-like pattern the
numbers~$\Lambda^{\varepsilon}_{n}$ follow (for any value of even~$h$) one
of the following is \Remph{always} the case:
\begin{align}
	&\Lambda^{\varepsilon}_{n}=\Lambda^{\varepsilon}_{n-2}
	=\Lambda^{\varepsilon-1}_{n-1}=\Lambda^{\varepsilon+1}_{n-1}=m		\M;\tag{i} \\
	&\Lambda^{\varepsilon}_{n}=\Lambda^{\varepsilon-1}_{n-1}=m, \quad
	\Lambda^{\varepsilon+1}_{n-1}=\Lambda^{\varepsilon}_{n-2}=m-1		\M;\tag{ii} \\
	&\Lambda^{\varepsilon}_{n}=\Lambda^{\varepsilon+1}_{n-1}=m, \quad
	\Lambda^{\varepsilon-1}_{n-1}=\Lambda^{\varepsilon}_{n-2}=m-1		\M;\tag{iii}\\
	&\Lambda^{\varepsilon}_{n}=m, \quad \Lambda^{\varepsilon}_{n-2}=m-2, \quad
	\Lambda^{\varepsilon-1}_{n-1}=\Lambda^{\varepsilon+1}_{n-1}=m-1		\M;\tag{iv}
\end{align}
with $m$ some positive integer and in each such case, the identity~\eqref{qL1T2} holds.

\thmpara
This concludes the proof.
\end{Proof}

\begin{Theorem}\ordlabel{TBigTheorem}
The coefficients of the Virasoro characters in the expansion~\eqref{qCharCoef} for all the
affine models are given
by the Chebychev polynomials of equation~\eqref{qZIsCheb}. In other words
\begin{equation}
	\sum_{\substack{\mu\in\expts{\G}\\\varepsilon\in\TaA_{[2h]}}}
	\frac{\phi^{(2\mu)}_{\varepsilon}}{\phi^{(2\mu)}_0}\,
	\psi^{(\mu)\ast}_a\,\psi^{(\mu)}_b\,
	\Lambda^{\varepsilon}_{n}
	=
	\chebychev[ab]{n}{\G}
	\M.
\mathlabel{qT1T1}
\end{equation}
Here $\chebychev{n}{\G}$ denotes the Chebychev polynomial of the
second kind satisfying: $\chebychev{0}{\G}=\II$ and $\chebychev{1}{\G}=\G$
and $\Lambda^{\varepsilon}_{n}$ again denotes the coefficient
$\Lambda^{\Nmodel{\TaA_{[2h]}}{0}{\varepsilon}}_{n}$.
\glick\end{Theorem}
\begin{Proof}
	This follows by induction on~$n$ using \saylemma~\ref{TTennessee}, the eigenvalue property
	\begin{equation}
		\beta^{(2\mu)}_{[2h]}=\beta^{(\mu)}_{[h]}
		\M,
	\end{equation}
	and the identity
	\begin{equation}
		\sum_{\lambda}\TaA_{[h]\varepsilon\lambda}\,F(\lambda)=F(\,\varepsilon-1 \!\!\!\!\pmod{h}\,)
			+F(\,\varepsilon+1 \!\!\!\!\pmod{h}\,)
		\M.
	\end{equation}
\end{Proof}

Thus we arrive at the following general form for the partition functions of each of
the affine Pasquier models on the cylinder:
\begin{equation}
%\Rboxed{
	\Tpart^{\Nmodel{\G}{a}{b}}(q)
	=
	\sum_{n=0}^{\infty}
	\chi_{\frac{n^2}{4}}(q)
	\,
	\chebychev[ab]{n}{\G}
%}
	\M.
\mathlabel{qZIsCheb}
\end{equation}

%%%%%%%%%%
%
\Section{The Role of Coxeter Geometry\ordlabel{sTROCG}}
We now establish the role played in the partition functions by the Steinberg-ordered
affine Coxeter element defined in section~\ref{sIAM} (equation~\eqref{qCoxChoice}).
Given the definition of this Coxeter element, we exclude the models based on the odd cycles~$\TaA_{[2h-1]}$
from our analysis.

The Coxeter elements associated with any classical algebra are known to be mutually conjugate and
therefore possess identical spectra~\cite{bHumphreys}.
The question of conjugacy of the affine Coxeter elements was addressed by
Berman \etal~\cite{pBLM89} and also by Coleman~\cite{pCol89}.
In the case of the affine tree graphs, \ie the~$\TaD$- and $\TaE$-cases,
it can be proved that all Coxeter elements are conjugate (see also~\cite{bHumphreys}).
Since there is no special choice of Coxeter element, we choose the most convenient example to manipulate.
However, in the
case of the cycles, the $\TaA$-graphs, the Coxeter elements do fall into a number of distinct spectral classes.
Indeed,
the Coxeter elements associated with the graph~$\TaA_{n-1}$ fall into~$[n/2]$ spectral classes.
In particular they do not belong all to a single conjugacy class.
Regardless, we
will find the affine generalisation of the Steinberg-ordered Coxeter element to be the most useful even in this
case. It possesses the property of being a representative of the largest spectral class. When~$n$ is even, this
largest spectral class is unique and might be taken to
define a natural choice of sorts
%(recall that, coincidentally, we ignore the odd cycles~$\TaA_{2n}$ in this paper),
although we do not make use of this property.
In any case, we remark that any Coxeter element and its inverse, both
belong to the same spectral class.

Let~$\kmalg$ denote an affine Kac-Moody algebra with Coxeter-Dynkin diagram~$\G$ (with~$l+1$ nodes).
Let~$\liealg$ denote its horizontal subalgebra.
We denote by~$\Pi\equiv\Rset{\rootA[a]}{a\in 0,1,\ldots l}$ the set of simple roots
with~$\rootA[0]$ the affine root and by~$\Rset{\weightA[a]}{a\in 0,1,\ldots l}$ the basis of
fundamental weights. We denote by~$\reflection{a}$ the Weyl-reflection defined by~$\rootA[a]$
so that
\begin{equation}
	\lambdaA\mapsto\reflection{\rootA}\lambdaA
	\definedas
	\lambdaA-\ablf{\lambdaA}{\corootA}\rootA
	\M,
\mathlabel{qAffineReflect}
\end{equation}
where~$\ablf{\:}{}$ denotes the inner (scalar) product within~$\rootSPA$
and~$\corootA\equiv 2\,\rootA/\!\ablf{\rootA}{\rootA}$ the coroot of~$\rootA$.
The behaviour of the weights under Weyl reflections
is given by
\begin{equation}
	\reflection{a}\weightA[b]
	=
	\weightA[b]-\delta_{ab}\,\rootA[b]
\mathlabel{qWeylWeight}
\end{equation}
as the weights and the coroots are dual.
We recall (see~\cite{bFuchs} for example) that the roots of an affine Kac-Moody algebra may be written in the
form~$\kmvec{\horiz{\alpha}}{k}{d}$ where~$\horiz{\alpha}$ is a root of~$\liealg$ and~$k$ and~$d$ are
respectively the components \wrt the centre and derivation of~$\kmalg$.
Denote the roots of $\kmalg$ by $\rootSA$.
A subset of~$\rootSA$ may be identified with the root system~$\rootSB$ of~$\liealg$ by
\begin{equation}
	\kmvec{\rootB}{0}{0}\equiv \rootB
	\M.
\end{equation}
The positive roots~$\rootSA[+]$ of~$\kmalg$ are
\begin{equation}
	\rootSA[+]
	\definedas
	\Rset[\big]{
	\rootA = \kmvec{\rootB}{0}{n}
	\in \rootSA
	}{
	\text{ $n>0$ or $n=0$, $\rootB\in\rootSB[+]$}
	}
	\M,
\mathlabel{qPosRoots}
\end{equation}
and the negative roots $\rootSA[-]\equiv\rootSA\smallsetminus
\rootSA[+]$.
We use the notation $\rootA>0$ for positive roots and
$\rootA<0$ for negative roots.
Given the choice~\eqref{qPosRoots}, simple roots exist and are
\begin{equation}
	\rootA[i]
	=
	\kmvec{\rootB[i]}{0}{0}
	=
	\rootB[i]
	\quad
	\text{for $i=1,\ldots,l$}
	\M;
\end{equation}
where~$\KosPiB\equiv\Rset{\rootB[a]}{a\in\G\smallsetminus\{0\}}$
are the simple roots of $\rootSB$;
together with
\begin{equation}
	\rootA[0]
	=
	\kmvec{-\highroot}{0}{1}
	=
	\delta - \highroot
	\M.
\end{equation}
$\highroot$ is the highest root of~$\liealg$ and~$\delta\equiv\kmvec{0}{0}{1}$.
Note that~$\delta$ is an imaginary or lightlike root and is invariant under the action of~$\aWeyl$.

If we regard~$\G$ as an operator on the root space~$\rootSPA$ of~$\kmalg$, \ie
\begin{equation}
\begin{split}
	\G : \rootSPA &\rightarrow \rootSPA \M,\\
		\rootA[a] &\mapsto \sum_{b\in\G} \G_{ab}\,\rootA[b]
		\M.\\
\end{split}
\mathlabel{qGop}
\end{equation}
then it may be shown that
\begin{Lemma}[Kostant~\cite{pKos85}]
\ordlabel{TKos3p8}
	The action of~$\G$ regarded as an operator on~$\rootSPA$ is
	\begin{equation}
		\G = \coxAR[1] + \coxAR[2]
		\M;
	\mathlabel{qGtoGeom}
	\end{equation}
	where $\coxAR[1]$ and $\coxAR[2]$ are the involutions defined in section~\ref{sIAM}.
\glick\end{Lemma}
This~\eqref{qGtoGeom} may be substituted into~\eqref{qZIsCheb} to obtain a geometric expression for the partition functions.
Performing the calculation however, it is not immediately obvious what sort of relationship we actually have.
Instead we proceed in a different way:

Choose the following as orbit representatives for the Steinberg-ordered Coxeter element~$\coxA$ defined in
section~\ref{sIAM}:
\begin{equation}
	\orbA[a]\definedas (1-\coxA^{-1})\weightA[a]
	\M,
\mathlabel{qOrbitWeight}
\end{equation}
or equivalently, by~\eqref{qWeylWeight} and~\eqref{qCoxChoice},
\begin{equation}
	\orbA[a]=\begin{cases}
		\rootA[a]       & \text{; if $\rootA[a]\in\Pi_1$.} \\
		\coxAo\rootA[a] & \text{; if $\rootA[a]\in\Pi_2$.}
	\end{cases}
	\M.
\end{equation}
This choice is a generalisation of the representatives chosen by Kostant~\cite{pKos59} for the orbit representatives of
the non-affine (Steinberg-ordered) Coxeter element. As it is a generalisation it is easy to see that this set is
also linearly independent.
Furthermore, their orbits are distinct; indeed
if one supposes~\cite{lDor92} that~$\orbA[b]$ lies in the Coxeter orbit of another:
\ie $\coxA^p \orbA[a]=\orbA[b]$ for some values of~$a$, $b$ and~$p$; then,
by \eqref{qOrbitWeight},
\begin{equation}
	\coxA^p \weightA[a] = \weightA[b]
	\M.
\end{equation}
This cannot be true as all fundamental weights are dominant
highest weights and are therefore not related to each other by
any Weyl element~\cite{bHumphreys2}.
The representatives~\eqref{qOrbitWeight} are uniquely characterised as the
positive roots which go negative (\ie whose images are negative roots) under
the action of the (affine) Coxeter element~$\coxA$. 
Now
\begin{Proposition}
\ordlabel{TGAction}The action of the operator $\G$ (see equation~\eqref{qGop}) on the orbit
representatives~$\orbSA\equiv\Rset{\orbA[a]}{a\in 0,1,\ldots,l}$, is given by
	\begin{equation}
		\sum_{c\in\G} \orbA[c] \, \G_{cb} =
		\begin{cases}
			(1+\coxA^{-1})\,\orbA[b] & \text{; if $\rootA[b]\in\Pi_1$.} \\
			(1+\coxA)\,\orbA[b]      & \text{; if $\rootA[b]\in\Pi_2$.}
		\end{cases}
	\M.
	\end{equation}
\glick\end{Proposition}
\begin{Proof}
	The proof is by simple calculation using definition~\eqref{qOrbitWeight}, equation~\eqref{qWeylWeight}
	and \saylemma~\ref{TKos3p8}.
\end{Proof}

In analogy with~\cite{pDor93}, we introduce the function
\begin{equation}
	\dorey_a\equiv\begin{cases}
			0 & \text{; if $\rootA[a]\in\Pi_2$.} \\
			1 & \text{; if $\rootA[a]\in\Pi_1$.}
		\end{cases}\M;
\mathlabel{qDoreyFn}
\end{equation}
and define $\dorey_{ab}\equiv\dorey_a-\dorey_b$.

We now demonstrate the connection between the Chebychev polynomials
and the affine Coxeter element. Dorey~\cite{pDor93} demonstrated
using explicit forms for the eigenvectors in the intertwiners~\eqref{qSocInt} that the partition functions of
the \Remph{classical} Pasquier models could be expressed as
\begin{equation}
	\Tpart^{\Nmodel{\G}{a}{b}}(q)
	=
	\sum_{\coxB^{-p}\orbB[b]\in\rootSB[+]}
	\chi_{1,2p+1+\dorey_{ab}}^c(q)
	\,
	\blf{\weightB[a]}{\coxB^{-p}\coorbB[b]}
	\M.
\mathlabel{qDoreyZ}
\end{equation}
The quantities $\weightB[a]$, $\coxB$ and $\orbB[b]$ refer to the root system and Weyl group
of a classical Lie algebra~$\liealg$ with adjacency matrix $\G$ and Coxeter number~$\horiz{h}$;
the scalar product
$\blf{}{\!}$ is the usual Euclidean bilinear form and the Virasoro characters
$\chi_{a,b}^c(q)$ refer to representations of the conformal algebra
with central charge $c=1-\frac{6}{\horiz{h}(\horiz{h}-1)}<1$.
As remarked upon earlier (demonstrated in the affine case), intertwiners and the Chebychev polynomials
are related. Thus we are led to conjecture a similar form for the
affine partition functions by equating the natural affine
generalisation of the inner product appearing in~\eqref{qDoreyZ} to the
Chebychev polynomials of the affine adjacency matrices.

\begin{Theorem}
\ordlabel{TGeometry}
	Let $\dorey_{ab}$ be defined as above~\eqref{qDoreyFn}. If $\G$ is the Coxeter-Dynkin
	diagram of the Kac-Moody algebra $\kmalg$ and $\ablf{}{\!}$ the \Remph{affine}
	bilinear form (scalar product) on the maximal Cartan subalgebra $\weightSPA$ 
	of $\kmalg$; then
	\begin{equation}
		\chebychev[ab]{2p+\dorey_{ab}}{\G}
		=
		\ablf{\weightA[a]}{\coxA^{-p}\coorbA[b]}
		\M;
	\mathlabel{qChebIsGeo}
	\end{equation}
	$\coxA$ denotes the (Steinberg-ordered affine) Coxeter element defined in
	equation~\eqref{qCoxChoice}.
\glick\end{Theorem}
%
%\thmpara
\begin{Proof}
The proof is by induction on the variable $N$ in $\chebychev{N}{\G}$.
The proof splits into four subproofs corresponding to the four different ways of assigning the
roots~$\rootA[a]$ and~$\rootA[b]$ amongst the sets~$\Pi_1$ and~$\Pi_2$.
In each case we first check that the form $\ablf{\weightA[a]}{\coxA^{-p}\coorbA[b]}$ reproduces
the correct initial conditions~\eqref{qChebRecurs}, \ie~$\chebychev{0}{\G}=\II$ and~$\chebychev{1}{\G}=\G$.
This is done using the duality between the weights and the (co-)roots and the fact that the affine
scalar product~$\ablf{\:}{}$ is normal, \ie
\begin{equation}
	\ablf{\eta \lambdaA}{\eta \lambdaA'}
	=
	\ablf{\lambdaA}{\lambdaA'}
	\quad\text{$\forall$ $\eta\in\aWeyl$; $\lambdaA,\lambdaA'\in\rootSPA$}
	\M.
\mathlabel{qNormal}
\end{equation}
The induction hypothesis in each case may be shown using repeated use
of proposition~\ref{TGAction}.
The full details of the proof may be found in~\cite{hTal}.
\end{Proof}

Thus we rewrite equation~\eqref{qZIsCheb} as
\begin{gather}
%\Rboxed{
	\Tpart^{\Nmodel{\G}{a}{b}}
	=
	\sum_{p=p'}^{\infty} \chi_{\frac{(2p+\dorey_{ab})^2}{4}}(q)
	\ablf{\weightA[a]}{\coxA^{-p}\coorbA[b]}
%}
\mathlabel{qGeomZ}
\intertext{with:}
	p'
	=
	\begin{cases}
		0 & \text{; if $\dorey_{ab}=0$ or $1$.} \\
		1 & \text{; if $\dorey_{ab}=-1$.}
	\end{cases}
	\M.
\end{gather}
This expresses the affine partition functions in terms of the
affine Coxeter element. Note that the summation automatically excludes
those terms which are zero due to the requirement that~$|a-b|$ and~$n$,
now replaced by~$2p+\dorey_{ab}$, be both even or both odd.

We remark that~\cite{pDor93} demonstrated the form~\eqref{qDoreyZ}
for the classical models by an entirely different method.
We duplicate this result using the method just described by recalling that the intertwiner
$V^{\lambda}_{ab}$ relating the $\G=D,E$ models to an $A$ model is equivalent
to a Chebychev polynomial by equation~\eqref{qdiFResult}.
%\begin{equation}
%	V^{\lambda}_{ab}=\chebychev[ab]{\lambda-1}{\G}
%\end{equation}
Theorem~\ref{TGeometry}
holds in the classical case also once one replaces the affine scalar product $\ablf{}{\!}$ with the
usual one $\blf{}{\!}$ and the affine Coxeter transformation, weights and orbit representatives with
their classical
analogues.
Thus we may equate the intertwiner with a geometric expression. The result~\eqref{qDoreyZ}
follows directly from~\cite{pSB89}
\begin{equation}
	\Tpart^{\Nmodel{\G}{a}{b}}(q)
	=
	\sum_{\lambda=1}^{h-1}
	\chi_{1,\lambda}^{c}(q)
	\,
	V^{\lambda}_{ab}
	\M.
\end{equation}

%%%%%%%%%%
%
\Section{Geometrical Consequences\ordlabel{sGC}}
In analogy with~\cite{pDor93}, we may attempt to construct invariant subspaces under the action of~$\coxA$
as follows:

Define
\begin{equation}
	\irootA[\mu]{i}\definedas
	\sum_{\substack{a\\\text{such that}\\\rootA[a]\in\Pi_i}} \psi_a^{(\mu)} \rootA[a]
	\quad\text{for $i=1,2$.}
\mathlabel{qInvntVecs}
\end{equation}
These vectors would appear to possess the properties:
\begin{equation}
\begin{split}
	\coxAR[i]\irootA[\mu]{i}
	&=
	-\irootA[\mu]{i}
		\M,\\
	\coxAR[j]\irootA[\mu]{i}
	&=
	\irootA[\mu]{i} + \beta^{(\mu)} \irootA[\mu]{j}
	\quad\text{for $i\neq j$.}
		\\
\end{split}
\end{equation}
So that $V^{(\mu)}\definedas\Rspan\{\irootA[\mu]{1},\irootA[\mu]{2}\}$ is closed.

Unfortunately, for some values of $\mu$ and $i$, such vectors~\eqref{qInvntVecs}
may be identically zero; this can be verified by explicit calculation.
However, for $\mu=0$ it is easy to see that
the properties of the Perron-Frobenius eigenvector ensure that
$\irootA{1}$ and $\irootA{2}$ are both non-zero, distinct and linearly independent.
%\pubnote{say why}
Hence the $\mu=0$ invariant subspace exists.
Choose the Perron-Frobenius eigenvector to
have all coefficients positive. We change notation and write
\begin{equation}
	\zroot[i]=
	\sum_{\substack{a\\\text{such that}\\\rootA[a]\in\Pi_i}} \psi_a^{(0)} \rootA[a]
	\M.
\end{equation}
So that
\begin{equation}
	\coxAR[i]\,\zroot[j]=\begin{cases}
		-\zroot[j]	       &\text{; if $i=j$.}	\\
		\zroot[j]+2\,\zroot[i] &\text{; if $i\neq j$.}
	\end{cases}
	\M.
\mathlabel{qCoxZroot}
\end{equation}
We also define the weights
\begin{equation}
	\zweight[i]=
	\sum_{\substack{a\\\text{such that}\\\rootA[a]\in\Pi_i}}  \psi_a^{(0)} \weightA[a]
	\M.
\end{equation}
Their action upon the $\{\zroot[i]\}$ is
\begin{equation}
	\ablf{ \zweight[i] }{ \cozroot[j] }
	=
	\half\,\delta^{ij}
	\M;
\mathlabel{qHalfDual}
\end{equation}
and upon roots in general, their action is to project onto the spaces $\Rspan\{\zroot[1]\}$ and $\Rspan\{\zroot[2]\}$
respectively. %\pubnote{need to be able to prove this!}
Equation~\eqref{qHalfDual} indicates that $2\,\zweight[i]$ is the dual
of $\co{\zroot[i]}$; therefore
the projection operator $\zproj$ onto the space $V\equiv\Rspan\{\zroot[1],\zroot[2]\}$
is given by
\begin{equation}
	\zproj \lambda \equiv
	2 \ablf{ \zweight[1] }{ \co{\lambda} } \zroot[1] +
	2 \ablf{ \zweight[2] }{ \co{\lambda} } \zroot[2]
	\M;
\mathlabel{qZprojDef}
\end{equation}
where $\lambda\equiv\kmvec{\horiz{\lambda}}{0}{n}$ is any root.
In particular $\zproj^2=\zproj$. We let $\zrootX\equiv\zroot[1] + \zroot[2]=\psi_0^{(0)}\delta$ and
$\zweightX\equiv\zweight[1] + \zweight[2]$ so that $\ablf{ \zweightX }{ \zrootX }=1$.

We remark that $\zweightX$ as defined, is a positive integer sum over the
weights $\Rset{ \weightA[a] }{ a\in\G }$. Using the fact that any
root may be written as a positive or negative integer sum of the simple
roots; we see that, by analogy with~\cite{pDor93},
\begin{equation}
\begin{split}
	&		   	\text{ $\lambda$ is a positive root}					\\
	& \Leftrightarrow	\ablf{ \weightA[a] }{ \co{\lambda} } > 0 \quad\text{for any $a$}		\\
%	& \Leftrightarrow	\ablf{ \weightA[a] }{ \co{\lambda} } \geq 0 \quad\text{for every $a$}	\\
	& \Leftrightarrow	\ablf{ \zweightX }{ \co{\lambda} } > 0					\M.\\
\end{split}
\mathlabel{qDoreyArg}
\end{equation}
Thus we have:
\begin{Lemma}\ordlabel{TDorey}
Any root of $\rootSA$ is a positive root, iff the scalar product of its coroot
with $\zweightX$ is positive; \ie
\begin{equation}
	\lambda\equiv\kmvec{\horiz{\lambda}}{0}{n}\in\rootSA[+]
	\quad\text{iff}\quad\ablf{\zweightX}{\co{\lambda}}>0
	\M.
\end{equation}
\glick\end{Lemma}

The following result is perhaps obvious; however we state it for clarity:
%
%%%\Newpage %FINAL
\begin{Proposition}
\ordlabel{Twjcommute}
	The projection $\zproj$ commutes with the Weyl group operators $\coxAR[1]$ and $\coxAR[2]$;
	and hence with the Coxeter element.
	That is
	\begin{equation}
		\restrictmap{\rule{0pt}{3ex}\commut{ \zproj }{ \coxAR[i] }\rule{0pt}{3ex}}{\rootSA}=0\M,
		\quad\text{for $i=1,2$}
	\end{equation}
	insofar as they act upon the roots $\rootSA$.
\glick\end{Proposition}
\begin{Proof}
	The simple roots $\KosPiA$ provide a basis for $\rootSA$. Therefore
	one need only examine to see if
	$\zproj\coxAR[i]\rootA[a]=\coxAR[i]\zproj\rootA[a]$ for each $a$ and $i$. This
	follows easily
	using the properties~\eqref{qCoxZroot} and the definition~\eqref{qZprojDef}.
\end{Proof}

For convenience, for any operator $\mathsf{Y}$, let
\begin{equation}
	\incase{b}{\mathsf{Y}}\definedas
	\begin{cases}
		\II        & \text{; if $\rootA[b]\in\Pi_1$.} \\
		\mathsf{Y} & \text{; if $\rootA[b]\in\Pi_2$.}
	\end{cases}
	\M.
\end{equation}
We also define
\begin{equation}
	\parity{a}\definedas\begin{cases}
		1 & \text{; if $\rootA[a]\in\Pi_1$.} \\
		2 & \text{; if $\rootA[a]\in\Pi_2$.}
	\end{cases}
	\M.
\end{equation}

\begin{Theorem}\ordlabel{TPositive}
	For all $a$, $b$ $\in\G$ and $p\in\ZZ^{+}$:
	\begin{equation}
		\ablf{ \weightA[a] }{ \coxA^{-p} \coorbA[b] } \geq 0
		\M.
	\end{equation}
\glick\end{Theorem}
\begin{Proof}
	By lemma~\ref{TDorey},
	we need only establish that $\ablf{ \zweightX }{ \coxA^{-p} \orbA[b] } >0$.

	\thmpara
	By proposition~\ref{Twjcommute}:
	\begin{equation}
	\begin{split}
		\zproj \coxA^{-p} \orbA[b] &=
		\coxA^{-p} \, \incase{b}{\coxAR[1]} \, \zproj \, \rootA[b] \\
		&=
		2\ablf{\zweight[1]}{\corootA[b]} \coxA^{-p} \, \incase{b}{\coxAR[1]} \, \zroot[1] +
		2\ablf{\zweight[2]}{\corootA[b]} \coxA^{-p} \, \incase{b}{\coxAR[1]} \, \zroot[2]
		\\&=
		2\ablf{\zweight[\parity{b}]}{\corootA[b]}\coxA^{-p}\,\incase{b}{\coxAR[1]}\,\zroot[\parity{b}]
%		\begin{cases}
%			2\ablf{\zweight[1]}{\rootA[b]}\coxA^{-p}\zroot[1] & \text{; $b\in\KosPA[1]$} \\
%			2\ablf{\zweight[2]}{\rootA[b]}\coxA^{-p}\coxAR[1]\zroot[2]
%									& \text{; $b\in\KosPA[2]$}
%		\end{cases}
		\M;\\
	\end{split}
	\end{equation}
	where we make use of the fact that $\ablf{\zweight[i]}{\corootA[b]}=0$ unless $\rootA[b]\in\Pi_i$.
	Thus we need only examine the Coxeter orbits of $\zroot[1]$ and
	$\coxAR[1]\,\zroot[2]\equiv2\,\zroot[1]+\zroot[2]$.

	\thmpara
	We see from~\eqref{qCoxZroot} that the application of $\coxA^{-1}$ to
	$(n)\,\zroot[1]+(n-1)\,\zroot[2]$ yields $(n+2)\,\zroot[1]+(n+1)\,\zroot[2]$;
	\ie $n$ is replaced by $n+2$. The roots of both orbits follow this form.
	Hence, as~$\zproj$ projects onto the invariant space~$V$,
	\begin{equation}
	\begin{split}
		\ablf{\zweightX}{\coxA^{-p}\coorbA[b]}
		&=
		\ablf{\zweightX}{\zproj\coxA^{-p}\coorbA[b]}
		\\&=
		\begin{cases}
			2\ablf{\zweight[1]}{\corootA[b]}(2\,p+1) & \text{; when $\rootA[b]\in\Pi_1$.} \\
			2\ablf{\zweight[2]}{\corootA[b]}(2\,p+3) & \text{; when $\rootA[b]\in\Pi_2$.}
		\end{cases}
		\M.\\
	\end{split}
	\end{equation}
	This may be rewritten as
	\begin{equation}
		\ablf{\zweightX}{\coxA^{-p}\coorbA[b]}=
		2\ablf{\zweightX}{\corootA[b]}\left(2\,p+1+2\,\delta_{2,\parity{b}}\right)
		\M.
	\end{equation}
	The scalar product appearing on the RHS clearly exceeds zero as $\rootA[b]$ is a
	simple root and hence a positive root. The remaining factor is positive for
	all $p\geq 0$.

	\thmpara
	This completes the proof.
\end{Proof}
Thus the coefficients~$\Lambda_n^{\Nmodel{\G}{a}{b}}$ appearing in the expression~\eqref{qCharCoef} are
all positive.
That they are integers follows
trivially from the duality of the weights and coroots together with the fact that
the set $\rootSA$ is closed under the action of~$\aWeyl$.

We mention as an aside that it might be interesting to investigate the relationship, if any, between the
projection~$\zproj$ and the defect map investigated by Berman \etal~\cite{pBLM89}.

Much of the structure of an affine algebra~$\kmalg$ is related
to the underlying structure of the horizontal subalgebra~$\liealg$.
If we separate out the translation part of the action from the affine Coxeter element we should be able to
see some of the structure of this subalgebra. Indeed, consider the translation~$\translat$ defined by
\begin{equation}
	\translat\definedas
	\reflection{0}\,\reflection{\highroot}
	\M;
\end{equation}
where~$\highroot$ is the highest root of~$\liealg$.
The action of $\translat$ on any $\lambdaA\equiv \kmvec{\lambdaB}{k}{d}$ is
\begin{equation}
	\translat\,\lambdaA=
	\kmvec{ \lambdaB-k\,\co{\highroot} }{ k }{
		d-\blf{\smash{\lambdaB}}{\smash{\co{\highroot}}}
		-\frac{2}{\blf{\highroot}{\highroot}}k }
	\M;
\end{equation}
and in particular, for any root $\rootA\equiv \kmvec{\rootB}{0}{n}$,
\begin{equation}
\begin{split}
	\translat\,\rootA
	&=
	\kmvec{ \rootB }{ 0 }{ n-\blf{\smash{\rootB}}{\smash{\co{\highroot}}} }	\\
	&=
	\rootA - \blf{\smash{\rootB}}{\smash{\co{\highroot}}} \delta	\M;\\
\end{split}
\mathlabel{qTranslat}
\end{equation}
\ie it `translates' any real root by $-\blf{\smash{\rootB}}{\smash{\co{\highroot}}}$
in the lightlike direction.
As $\reflection{0}=\translat\,\reflection{\highroot}$ we may regard
$\aWeyl$ as being generated by the set
\begin{equation}
	\left\{ \translat \right\} \cup
	\Rset{ \reflection{a} }{ a=1,\ldots,l }
	\M.
\end{equation}
$\aWeyl$ is the smallest group containing both the (classical) Weyl group~$\Weyl$
of~$\liealg$ and the affine translation~$\translat$.
Define $\coxEt$ by the requirement
\begin{equation}
	\coxAt=\translat\,\coxEt=\coxEt\,\itranslat
	\M,
\end{equation}
so that
$\coxEt$ is $\coxAt$ with the reflection $\reflection{0}$ replaced with
$\reflection{\highroot}$.
Define
\begin{equation}
\begin{split}
	&\coxEo\definedas\coxAo		\\
	\preword{and}&\coxE\definedas\coxEt\,\coxEo	\M;\\
\end{split}
\end{equation}
so that $\coxA\equiv\translat\,\coxE$. Note in particular
that $\coxE$, $\coxEo$ and $\coxEt$ are all elements of the Weyl group
$\Weyl$ of the horizontal subalgebra $\liealg$ and are therefore constrained
to be of \Remph{finite} order and to act trivially on the lightlike components of the
roots of~$\kmalg$.

We may use this ``Euclidean Coxeter element''~$\coxE$ to examine separately
the action of~$\coxA$ on the horizontal and lightlike parts of the
orbit representatives.
Consider
\begin{equation}
\begin{split}
	\coxA^{-1} \orbA[b]
	&= \coxA^{-1} \kmvec{\orbB[b]}{0}{\delta_{0b}} \\
	&= \coxE^{-1}\translat\kmvec{\orbB[b]}{0}{\delta_{0b}} \\
	&= \coxE^{-1}\kmvec{\orbB[b]}{0}{\delta_{0b} +
		\blf{\orbB[b]}{\co{\highroot}} } \\
	&= \kmvec{\coxE^{-1}\orbB[b]}{0}{\delta_{0b} +
		\blf{\orbB[b]}{\co{\highroot}}}	\M.\\
\end{split}
\end{equation}
Iterating, so that for $p\geq 0$,
\begin{equation}
	\coxA^{-p}\orbA[b] =
	\kmvec{\coxE^{-p}\orbB[b]}{0}{\delta_{0b}+
		\sum_{p'=0}^{p-1}\blf{\coxE^{-p'}\orbB[b]}{\co{\highroot}}}
	\M.
\end{equation}
As~$\coxE$ has period~$h$ (this is the definition of the affine Coxeter number, see~\cite{pSte85}
and~\cite{pBLM89}), we may write
\begin{equation}
	\coxA^{-(qh+r)}\orbA[b]=
	\kmvec{ \coxE^{-r}\orbB[b] }{ 0 }{ \delta_{0b} +
		\sum_{r'=0}^{r-1} \blf{ \coxE^{-r'}\orbB[b] }{ \co{\highroot} } +
		q \sum_{r'=0}^{h-1} \blf{ \coxE^{-r'}\orbB[b] }{ \co{\highroot} } }
	\M;
\mathlabel{qFinite}
\end{equation}
where~$p=qh+r$ and~$r<h$ by the usual division algorithm. Thus to calculate the
coefficient~$\ablf{ \weightA[a] }{ \coxA^{-p}\orbA[b] }$ we need only
examine the behaviour of the ``\Remph{Euclidean projection}'' of the Coxeter orbits
onto~$\rootSB$, the root system of $\liealg$. In particular, due to the
periodicity of~$\coxE$, we find that one can calculate any coefficient
from the knowledge of the~$h(l+1)$ vectors
\begin{equation}
	\Rset[\Big]{ \coxE^{-r} \orbB[b]
		}{ \text{ $r\in\ZZ$, $0\leq r\leq h-1$; $b\in\G$ } }
	\M,
\end{equation}
together with their inner products with the (dual of the) highest root
$\co{\highroot}$; in particular, the sums
\begin{equation}
	\Rset[\Big]{
	\Delta^{(r)}_b \definedas
	\sum_{r'=0}^{r-1}\blf{ \coxE^{-r'}\orbB[b] }{ \co{\highroot} }
	}{ \text{$r\in\ZZ$, $0 \leq r \leq h-1$} }
%	\M.
\end{equation}
determine the action of the translation~$\translat$ which is localised to the lightlike components
of~$\coxA^{-p}\orbA[b]$.
The action of~$\coxA$ is thus periodic on the horizontal components of the roots but translates
the lightlike components.
Note that it is the orbits $\gengroup{\coxE}\orbB[b]$ of the Euclidean Coxeter
element which appear in these quantities and \Remph{not} the orbits of the
Coxeter element $\coxB$ of the algebra $\liealg$. However it is interesting to
see that the \Remph{entire} structure of the affine Coxeter orbits can be determined
by examining the orbits of $\coxE$ alone. These being of finite cardinality, allow
one to determine the entire partition function~\eqref{qGeomZ} from a finite number of
calculations. Indeed this method of calculation was checked using Mathematica. The correct partition functions
are obtained.%

%%%%%%%%%%
%
\Section{Concluding Remarks\ordlabel{sCR}}
We have shown that the partition functions of the ($\ZZ_2$-colourable)
affine Pasquier models may be expressed in terms of the
orbit structure of the affine Coxeter element (see equation~\eqref{qGeomZ}).
We have thus generalised the geometric expression of~\cite{pDor93} from~$c<1$ conformal models
to~$c=1$.

In particular we have found expressions, \eqref{qZIsCheb} and~\eqref{qGeomZ} (the former valid for all
the models),
for the partition functions of the affine Pasquier models with fixed boundary conditions
on the cylinder which until now had been missing from the
literature. The interested reader may like to verify that, in the specific case of the
$4$-state Potts model with free boundary conditions, the partition function
implied by~\eqref{qZIsCheb} or~\eqref{qGeomZ} agrees with that found in~\cite{pSB89}.

The geometric expression~\eqref{qGeomZ}, whilst unlike~\eqref{qZIsCheb} is
not manifestly symmetric in the boundary
conditions, permits an easy verification of the positive integrality of the coefficients~$\Lambda^{\Nmodel{\G}{a}{b}}$
in equation~\eqref{qCharCoef}. 
%
%We have also observed, through this geometric expression, that the coefficients appearing within
%the partition functions~\eqref{qCharCoef} are positive integers.
%
%It is particularly interesting to note
We are also able to observe
that the physics of these models is expressible in terms of the physics
of the corresponding non-affine models albeit in a non-trivial way.
Indeed the quantities~$\coxE^{-r}\orbB[b]$ and~$\Delta^{(r)}_b$ are determined wholly within the structure
of the Lie algebra~$\liealg$ without reference to the peculiarities of the greater Kac-Moody algebra~$\kmalg$.
These quantities provide one with an exceptionally quick (\ie finite) method of evaluating the
partition function of any affine model. What initially appears to be an infinite sequence, may in
fact be reduced to a finite number of evaluations by equation~\eqref{qFinite}.
It might be possible that this can be used as a mechanism to collect the degenerate~$c=1$ Virasoro characters in the
partition functions in such a way as to observe the action of higher symmetries, such
as supersymmetry or those associated with $W$-algebras, in the physics of these models.
We remark that neither of these results are apparent from the purely algebraic expression~\eqref{qZIsCheb}.

Most importantly, the geometric result~\eqref{qGeomZ} and its consequences lend support to the idea that
the geometry of root systems has a more general role within conformal field theories and related structures.
This idea has been investigated for $S$-matrices by Dorey in~\cite{pDor} and~\cite{lDor92};
and for more general graph-lattice models by Zuber~\cite{pZub96}.
%(see also references~\cite{pDor}, \cite{lDor92} and~\cite{pZub96}).

%%%%%%%%%%
%
\Section{Acknowledgements\ordlabel{sA}}
This work is a subset of work carried out by the author between April~$1995$ and October~$1998$
in partial fulfillment of the requirements for the award of the
degree of Doctor of Philosophy of the University of Durham (England).
A full account of this and other work may be found elsewhere~\cite{hTal}.

The author wishes to express his thanks to Patrick Dorey for supervising this project
and for numerous valuable discussions.

%%%%%%%%%%
%

\end{document}